\documentclass[11pt]{article}
\usepackage{moriond,epsfig,subfigure,xspace}

\def\be{\begin{equation}}
\def\ee{\end{equation}}
\def\bea{\begin{eqnarray}}
\def\eea{\end{eqnarray}}

\newcommand{\invpb}{pb$^{-1}$}
\newcommand{\dzero}{D\O}

\newcommand{\met}{\mbox{$\not\!\!E_{{\rm T}}$}}

\newcommand{\ttbar}{\mbox{$t\bar{t}$}}
\newcommand{\emu}{\mbox{$e\mu$}}

\newcommand{\Ztt}{\mbox{$Z \to \tau\tau$}}
\newcommand{\etrack}{\mbox{$e+\rm{track}$}}
\newcommand{\mutrack}{\mbox{$\mu+\rm{track}$}}
\newcommand{\ltrack}{\mbox{lepton$+$track}}
\newcommand{\Zll}{\mbox{$Z \to \ell\ell$}}

\begin{document}
\vspace*{4cm}
\title{Measurement of the $\mathbf{t\bar{t}}$ Production Cross Section at $\mathbf{\sqrt{s}}$~=~1.96~TeV in the
Combined Lepton+Track and $e\mu$ Channel Using 370~pb$^{\mathbf{-1}}$ of \dzero\ Data}

\author{Sara Lager}

\address{Stockholm University, Department of Physics, \\
  106 91 Stockholm, Sweden}

\maketitle\abstracts{
A measurement of the $\ttbar$ production cross section at $\sqrt{s} =  1.96$ TeV in the 
dilepton final states using a lepton+track selection and secondary vertex $b$-tagging is presented.
One of the two leptons from the decay of the \ttbar\ pair is
allowed to be identified only as an isolated track to improve the selection efficiency. 
The result is combined with a measurement in the $\ttbar \to \emu$ final state.
The measurements are based on 370~\invpb\ of data collected with the \dzero\ experiment
at the Tevatron collider.
The preliminary cross section obtained in the combined lepton+track and \emu\ channel is:
$$
\sigma_{t\bar{t}} = 8.6 ^{+1.9}_{-1.7} \ ({\rm stat}) \pm 1.1 \ ({\rm syst}) \pm 0.6 \ ({\rm lumi})\ \rm{pb}.
$$ 
}

\section{Introduction \label{sec:intro}}

The top quark is the heaviest of the known fundamental fermions. Measurements of its production rate and properties allow for
precision tests of the theory of strong interactions, quantum chromodynamics (QCD), and deviations from the predictions may indicate
physics beyond the standard model (SM).

At the Tevatron, the dominant production mode for top quarks is $\ttbar$ pair production.
Theoretical calculations predict the \ttbar\ cross section ($\sigma_{t\bar{t}}$) with 
an uncertainty of less than 15\% \cite{top1,top2,top3}.
The top quark decays with almost 100\% probability into a $W$~boson and a $b$~quark.
The \ttbar\ final state is therefore completely determined by the decays of the two $W$~bosons. 
This report describes a measurement by the \dzero\ collaboration of $\sigma_{t\bar{t}}$
in $p\bar{p}$ collisions at $\sqrt{s} =  1.96$ TeV at the Tevatron collider.
The measurement is performed in the ``dilepton'' final states where both $W$ bosons decay leptonically.
It uses a ``\ltrack'' selection, where one lepton is required to be fully reconstructed whereas the second lepton is identified 
as an isolated track only. This selection improves the statistical sensitivity over one where both leptons are required to be fully
reconstructed. 

The data sample, collected with the \dzero\ detector, corresponds to an integrated luminosity
of approximately 370~\invpb. 
The measurement in the \ltrack\ channel is combined with a measurement in the \emu\ channel~\cite{emusummer} using the same data sample.

\section{Selection Criteria \label{sec:presel}}

The final state in the \ltrack\ channel is characterized by a high $p_T$ isolated lepton 
(electron or muon), a high $p_T$ isolated track, large missing transverse energy (\met) and two jets from $b$~quarks. 
The transverse momentum selections used are: electron, muon and track $p_T > 15$~GeV and jet $p_T > 20$~GeV. 
The \met\ selections ranges from 15 to 35~GeV depending on the reconstructed lepton flavor and the invariant mass of the lepton and track pair.
At least one of the jets is required to be identified as coming from a $b$ quark.

The $b$-tagging algorithm used is the secondary vertex tagging algorithm (SVT), 
which explicitly reconstructs vertices that are displaced from the primary vertex. 
The performance of the algorithm has been extensively studied in data. 
A more detailed description of the algorithm and its performance can be found elsewhere~\cite{ljets_plb}.

The $\ttbar \to \emu$ events are explicitly vetoed in the \ltrack\ analysis. 
These events do not suffer from $Z \to ee,\mu\mu$ backgrounds and the use of $b$-tagging 
to enhance the sample purity is therefore unnecessary.
The veto allows for a straightforward combination of the \ltrack\ measurement and 
a measurement in the \emu\ channel using only kinematical selections.

\section{Signal Acceptance and Background Estimates \label{sec:estimates}}

The preselection efficiency for \ttbar\ events is estimated using events generated with {\sc alpgen}~\cite{alpgen}, interfaced to 
{\sc pythia}~\cite{pythia} to provide higher order QCD evolution and short lived particle decays. 
Both $W$ bosons are forced to decay into a lepton-neutrino pair, including all $\tau$ final states.
The tagging probability is estimated by taking the jet kinematics from the \ttbar\ Monte Carlo sample and folding in
the per jet tagging efficiency parameterizations determined on data. 

The largest background in the lepton+track analysis comes from $Z\to ee,\mu\mu$ events with fake \met.
Additional backgrounds arise from multijet and $W$+jets events where the isolated lepton, the isolated track or 
both are the result of misreconstruction.
The irreducible backgrounds ($WW$ and \Ztt) are expected to be small.

The event kinematics of the irreducible backgrounds and the fake \met\ background  
are modelled using Monte Carlo samples generated with {\sc alpgen} 
interfaced to {\sc pythia}~(\Zll) or with {\sc Pythia}~($WW$).
The number of \Zll\ events is normalized using the number of observed events 
in the low \met\ region.
For the $WW$ process, the NLO cross section is used~\cite{WWnlo}.
The number of multijet and $W$+jets background events are estimated from data using a four-step matrix method~\cite{ltrack_conf}. 
The tagging probability for $W$+jets (multijet) events is estimated in a sample with an isolated lepton and
high (low) \met.
The $b$-tagging efficiency for $WW$ events is assumed to be the same as for $W$+jets events.

\section{Sample Composition After $b$-tagging \label{sec:sample_comp}}

\begin{figure}[btp]
  \begin{center}
    \subfigure[Lepton+track events as a function of the number of jets.
    \label{fig:final_pred}]{
      \includegraphics[width=0.30\textwidth]{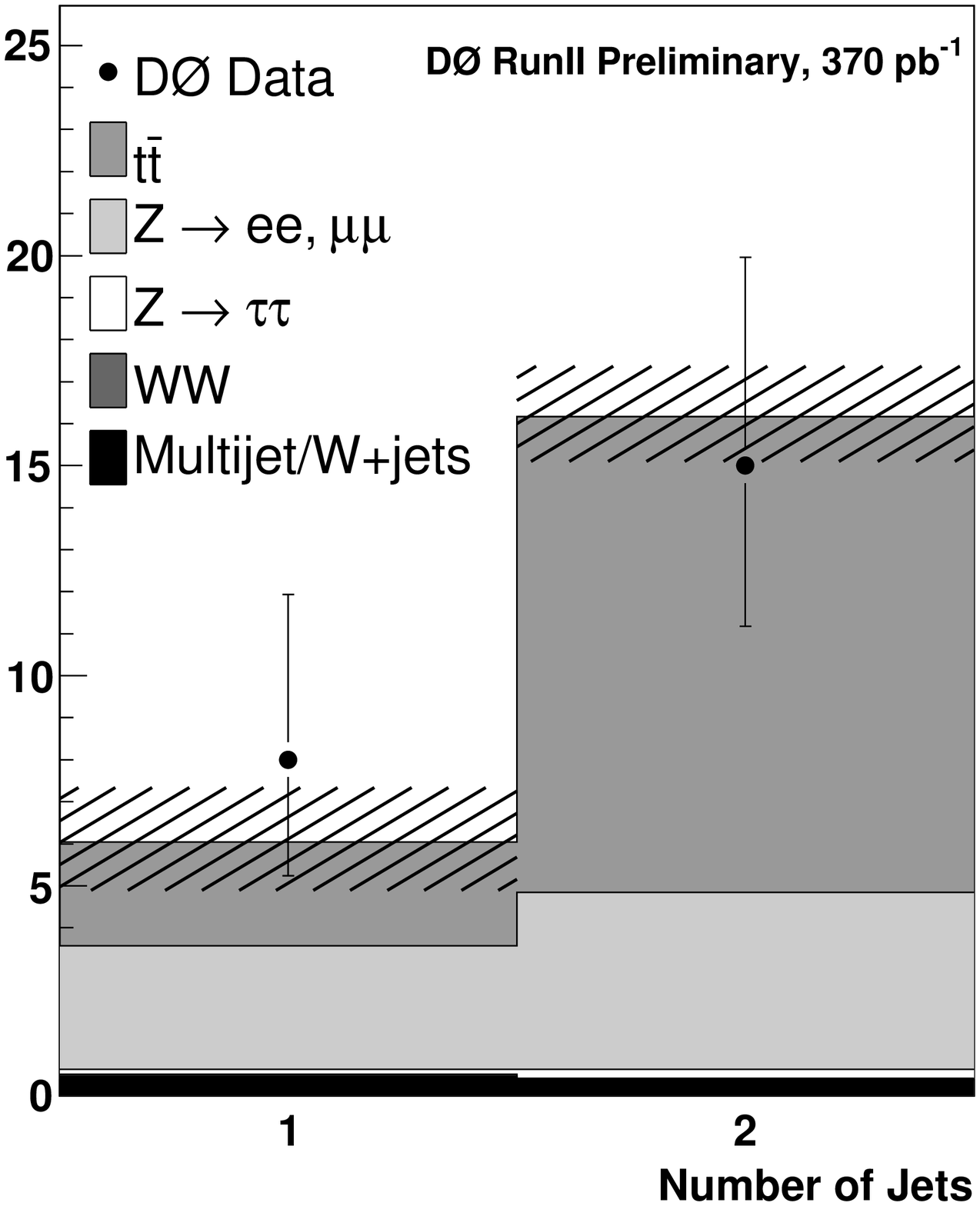}}
    \subfigure[Lepton+track and \emu\ events with $\geq$ 2 jets as a function of the \met.
    \label{fig:MET}]{
      \includegraphics[width=0.30\textwidth]{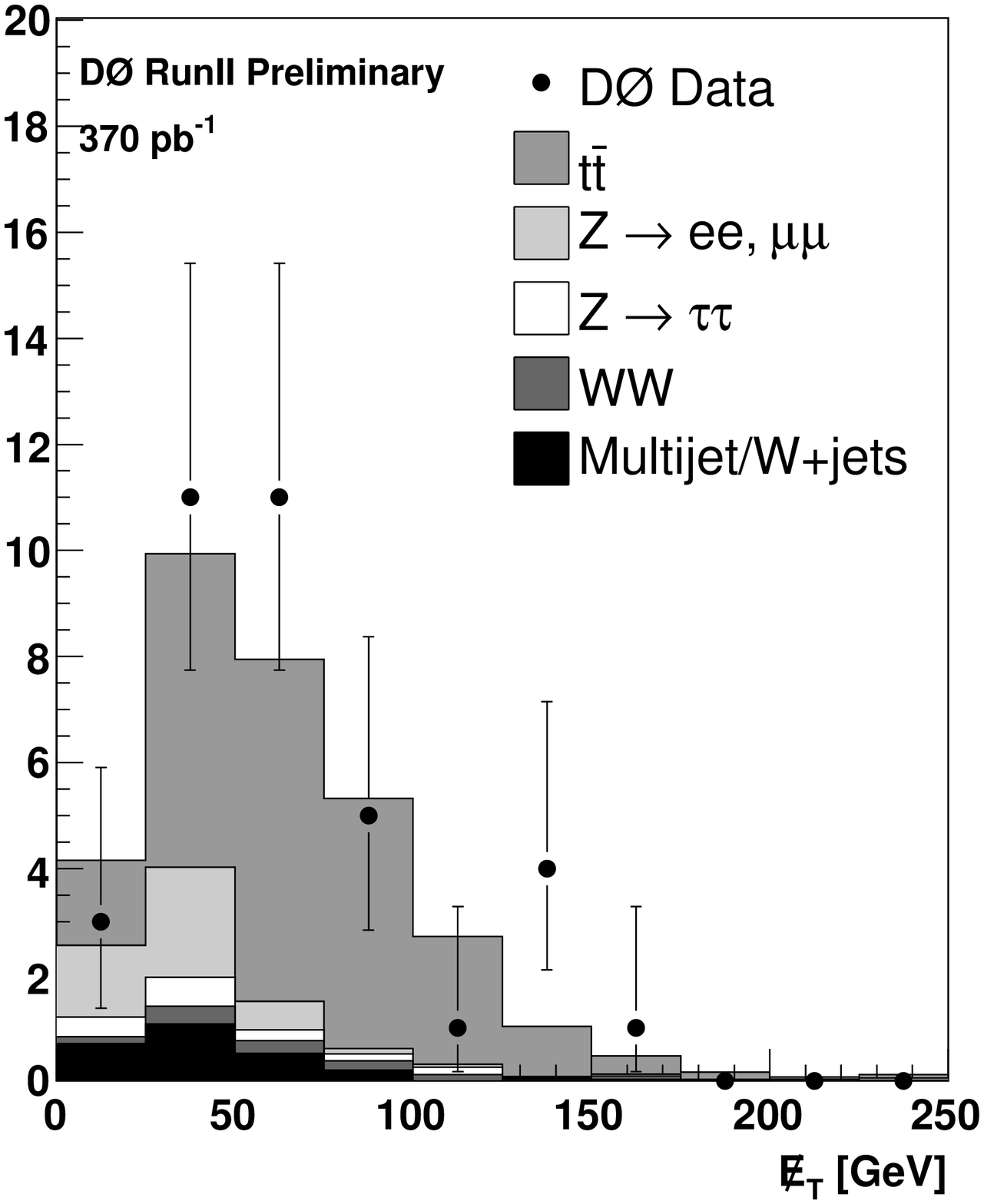}}
    \caption{The number of observed (points) and predicted (histogram) events in the lepton+track sample (a) and the 
      combined lepton+track and \emu\ sample (b). 
      The hashed area shows the total uncertainty on the prediction.
    }
  \end{center}
\end{figure}

Table~\ref{tab:master_tagged} shows the number of expected and observed events after $b$-tagging.
The expected number of \ttbar\ events is derived assuming $\sigma_{t\bar{t}} = 7$~pb.
The observed and predicted number of $b$-tagged events is also shown in Fig.~\ref{fig:final_pred}.
Figure~\ref{fig:MET} shows the agreement between the observed and predicted \met\ spectrum in
the combined lepton+track and \emu\ channel for events with two or more jets.
\begin{table}[htb]
\caption{Number of observed and predicted events after $b$-tagging.
The ``Total bkg prediction'' and the ``Total prediction'' include systematic uncertainties.
  \label{tab:master_tagged}}
\vspace*{0.4cm}
\begin{center}
\begin{tabular}{||l|cc|cc||}
\hline\hline
& \multicolumn{2}{|c|}{\bf{\etrack}}& \multicolumn{2}{|c|}{\bf{\mutrack}}\\
\hline
& Njets = 1 & Njets $\geq$ 2 & Njets = 1 & Njets $\geq$ 2 \\
\hline
$WW$ & 0.037 $\pm$ 0.002 & 0.010 $\pm$ 0.002 & 0.016 $\pm$ 0.001 & 0.009 $\pm$ 0.002 \\
$Z/\gamma^* \to \tau\tau$ & 0.09 $\pm$ 0.02 & 0.13 $\pm$ 0.02 & 0.03 $\pm$ 0.01 & 0.09 $\pm$ 0.02 \\
$Z/\gamma^* \to ee,\mu\mu$ & 1.49 $\pm$ 0.04 & 2.35 $\pm$ 0.06 & 1.44 $\pm$ 0.04 & 1.86 $\pm$ 0.06 \\
Multijet/$W$+jets & 0.36 $\pm$ 0.06 & 0.35 $\pm$ 0.07 & 0.08 $\pm$ 0.02 & 0.05 $\pm$ 0.03 \\
\hline 
\rule[0.22cm]{0pt}{1ex} 
Total bkg prediction & 1.97 $^{+0.91}_{-0.85}$ & 2.83 $^{+0.87}_{-0.64}$ & 1.57 $^{+0.77}_{-0.77}$ & 2.00 $^{+0.51}_{-0.49}$ \\
\hline
\ttbar & 1.55 $\pm$ 0.03 & 6.59 $\pm$ 0.07 & 0.92 $\pm$ 0.02 & 4.74 $\pm$ 0.06 \\
\hline
\rule[0.22cm]{0pt}{1ex} 
Total prediction & 3.53$^{+0.99}_{-0.86}$ & 9.4$^{+0.99}_{-0.85}$ & 2.49$^{+0.83}_{-0.77}$& 6.74$^{+0.67}_{-0.64}$ \\
\hline\hline
Data & 7 & 9 & 1 & 6 \\
\hline\hline
\end{tabular}
\end{center}
\end{table}

\section{Lepton+Track and $e\mu$ Channel Combination \label{sec:xsec}}

The combined cross section is estimated from the combination of five independent channels: $e\mu$,
$e$+track ($\mu$+track) with one jet and $e$+track ($\mu$+track) with two or more jets. The functional form of
the likelihood function for the $e\mu$ channel is described in \cite{emusummer}. 
There is no overlap between the \ltrack\ and \emu\ samples. 
The cross sections in the \ltrack\ and \emu\ channels, and the combined cross section are:
\begin{eqnarray*}
  \ltrack:\ \sigma_{t\bar{t}} & = & 7.1 \ ^{+2.6}_{-2.2} \ ({\rm stat}) \ ^{+1.3}_{-1.3} \ ({\rm syst}) \ \pm 0.5 \ ({\rm lumi})\ \rm{pb}. \\ 
  \emu:\ \sigma_{t\bar{t}} & = & 10.2 \ ^{+3.1}_{-2.6} \ ({\rm stat}) \ ^{+1.6}_{-1.3} \ ({\rm syst}) \ \pm 0.7 \ ({\rm lumi})\ \rm{pb}. \\ 
  {\rm comb}:\ \sigma_{t\bar{t}} & = & 8.6^{+1.9}_{-1.7}\:{\rm (stat)}\: ^{+1.1}_{-1.1}\:{\rm (syst)}\: \pm 0.6\:{\rm (lumi)}\:{\rm pb}.
\end{eqnarray*}
All cross sections are derived for $m_t=175\;\rm GeV$. 
In the top quark mass range of 170-180~GeV, the measured cross section changes by 0.085~pb per GeV.

The systematic uncertainty on the cross section measurement is obtained by varying the backgrounds and efficiencies,
within their errors, with all correlations between the channels and between the different classes of backgrounds taken 
into account.
Table \ref{tab:syscombo} summarizes the contributions
from the different sources of systematic uncertainties to the total systematic
uncertainty on the cross sections.

\begin{table}[tbp]
\begin{center}
\caption{The systematic uncertainties (in \%) on the \emu, \ltrack\ and combined result.
\label{tab:syscombo}}
\vspace*{0.4cm}
\begin{tabular}{||l||c|c||c|c||c|c||}
\hline \hline
Source & \multicolumn{2}{c||}{\emu} & \multicolumn{2}{c||}{l+track}& \multicolumn{2}{c||}{l+track+\emu}\\ \hline
Lepton selections                      & $-$6.6  & $+$7.4 & $-$5.7 & $+$6.8 & $-$5.6 & $+$6.2 \\ \hline
Jet identification                     & $-$3.5  & $+$4.7  & $-$2.5 & $+$0.1 & $-$3.2 & $+$2.2 \\ \hline
Jet energy calibration                 & $-$6.0  & $+$7.0 & $-$13.9 & $+$8.9 & $-$8.8 & $+$6.8 \\ \hline
Background modeling                    & $-$2.5  & $+$2.5 & $-$10.0 & $+$13.0 & $-$4.3 & $+$5.4 \\ \hline
$\ttbar$ tagging probability          & 0       & 0    & $-$3.9 & $+$4.1 & $-$2.1 & $+$2.2 \\ \hline
Trigger                                & $-$7.5  & $+$10.0 & $-$2.3 & $+$2.9 & $-$3.7 & $+$6.1 \\ \hline
Other                                  & $-$4.3  & $+$4.8 & $-$2.1 & $+$2.4 & $-$2.7 & $+$3.0 \\ \hline\hline
Total                                  & $-$12.4 & $+$16.1  & $-$18.9 & $+$18.5 & $-$12.8 & $+$13.3 \\ \hline
\hline 
\end{tabular}
\end{center}
\end{table}

\section{Summary \label{sec:summary}}

Using the combined lepton+track and \emu\ channel, the cross section for $\ttbar$ pair production in $p\bar{p}$ collisions 
at $\sqrt{s} = 1.96$ TeV is measured to be:
$
\sigma_{t\bar{t}}  =  8.6^{+1.9}_{-1.7}\:{\rm (stat)}\:  ^{+1.1}_{-1.1}\:{\rm (syst)}\: \pm 0.6\:{\rm (lumi)}\:{\rm pb}.
$
The data is in good agreement with the prediction from the standard model using
perturbative QCD calculations.

\section*{Acknowledgments}
We thank the staffs at Fermilab and collaborating institutions,
and acknowledge support from the
DOE and NSF (USA);
CEA and CNRS/IN2P3 (France);
FASI, Rosatom and RFBR (Russia);
CAPES, CNPq, FAPERJ, FAPESP and FUNDUNESP (Brazil);
DAE and DST (India);
Colciencias (Colombia);
CONACyT (Mexico);
KRF and KOSEF (Korea);
CONICET and UBACyT (Argentina);
FOM (The Netherlands);
PPARC (United Kingdom);
MSMT (Czech Republic);
CRC Program, CFI, NSERC and WestGrid Project (Canada);
BMBF and DFG (Germany);
SFI (Ireland);
The Swedish Research Council (Sweden);
Research Corporation;
Alexander von Humboldt Foundation;
and the Marie Curie Program.

\end{document}